# Detection of reactor antineutrino coherent scattering off nuclei with a two-phase noble gas detector

**Dmitri Akimov**[a*], **Alexander Bondar**[b], **Alexander Burenkov**[a], **and Alexei Buzulutskov**[b]

[a] *State Scientific Centre of Russian Federation Institute for Theoretical and Experimental Physics (ITEP), 25 Bolshaya Cheremushkinskaya, 117218 Moscow, Russia*
[b] *Budker Institute of Nuclear Physics (BINP), 11 Lavrentiev avenue, 630090 Novosibirsk, Russia*
  *E-mail*: akimov_d@itep.ru

ABSTRACT: Estimation of the signal amplitudes and counting rates for coherent scattering of reactor antineutrino off atomic nuclei in two-phase xenon and argon detectors has been done. A conceptual design of detector *based on the existing technologies* and experience has been proposed. It is shown that a condensed xenon/argon two-phase detector possesses the necessary sensitivity for the use in experiment on detection of coherent scattering of the reactor antineutrino off nuclei. It is shown that a two-phase detector with both optical readout by PMTs and ionisation readout by GEM/THGEM possesses superior capability for identification of the events of coherent antineutrino scattering.

---

[*] Corresponding author.

# Contents

## 1. Introduction

Coherent neutrino (antineutrino) scattering off an atomic nucleus

$$v + A \rightarrow v + A \tag{1}$$

is a fundamental process predicted by a Standard Model [1],[2],[3]. However, it has not been detected yet due to experimental difficulties: the energy deposition is in the energy range of ≤ 1 keV, and the detection mass must be significant, of an order of several or even more kilograms. Differential and total cross sections are described by the formulas:

$$\frac{d\sigma}{dT_A} = \frac{G_F^2}{4\pi} m_A [Z(1 - 4\sin^2\theta_W) - N]^2 \left[1 - \frac{m_A T_A}{2E_\nu^2}\right] \tag{2}$$

$$\sigma_{tot} = \frac{G_F^2 E_\nu^2}{4\pi} [Z(1 - 4\sin^2\theta_W) - N]^2 \tag{3}$$

where $m_A$ – nucleus mass, $T_A$ – kinetic energy of recoil nucleus, $E_\nu$ – neutrino energy, $Z$ – nucleus charge, $N$ – number of neutrons in the nucleus, $\sin^2\theta_w \approx 0.22$. Since the value $(1 - 4\sin^2\theta_w)$ is small and the total cross-section is then proportional to $N^2$, it is preferable to use the heavy elements as a target.

    Detectors of different types, capable to measure ionization in the energy region less then 1 keV were proposed and currently under development [4],[5],[6],[7],[8]. A two-phase detector based on liquid argon has been proposed in [7] to search for this process. The condensed xenon as a working medium even has not been considered there since the mean nuclear recoil energy released in the scattering of antineutrino off the xenon nucleus is by a factor of ~ 4 lower than that of argon. The estimations made in [6] for gaseous xenon and argon detectors also gave quite low signal values. However, recent experimental estimations of the yield of ionisation electrons from the track for the nuclear recoils in liquid Xe (LXe) performed by the Xenon10 Dark Matter search team [9] have shown unexpectedly high value of the ionization yield, up to

20 electrons per 1 keV of deposited energy when the data are extrapolated down to 1 keV. This will allow one to obtain the signal with a magnitude of up to several ionisation electrons per nuclear recoil.

In this work, an estimation of the signal amplitudes and counting rates for the coherent scattering of reactor antineutrino in a two-phase xenon detector has been made with the use the new data of ionisation yield for nuclear recoils in the liquid xenon and as well for the two-phase argon detector in assumption of the similar yield. A conceptual design of the detector *based on the existing technologies* and experience has been proposed. The aim of the current work is to show that a condensed xenon/argon two-phase detector possesses the necessary sensitivity for the use in experiment on detection of coherent scattering of the reactor antineutrino off atomic nuclei. Apart from fundamental issue, such a detector may found possible applications for monitoring of a nuclear reactor by measuring an antineutrino flux.

## 2. Signal estimation in liquid xenon and argon

Energy distribution of the nuclear recoils is obtained from the differential spectrum resulted by formula (2) convolved with the energy spectrum of antineutrino. An analytic expression for the spectrum can be taken from [10]:

$$\rho(E_v) = K \cdot \exp[a \cdot E_v + b \cdot E_v^2 + c \cdot (E_v/8)^{10}], \qquad (4)$$

were $E_v$ in MeV, $K$=5.09 (MeV·fission)$^{-1}$, a= -0.648, b= -0.0273, c= -1.41. The formula is obtained by fitting the experimental data in the range 2<$E_v$<9 MeV; it is normalized to the act of fission of a "standard composition" nuclear fuel. In evaluation of the energy deposition in the detector, this dependence was extrapolated to lower energies, down to the zero energy. Although the real antineutrino energy spectrum is different at the low energies, this does not affect the number of the events with nonzero charge because scattering of antineutrino with $E_\nu < 2$ MeV produces nuclear recoils with kinetic energies $T_A < 30$ eV, and thus, does not produce ionisation in the noble gas detector.

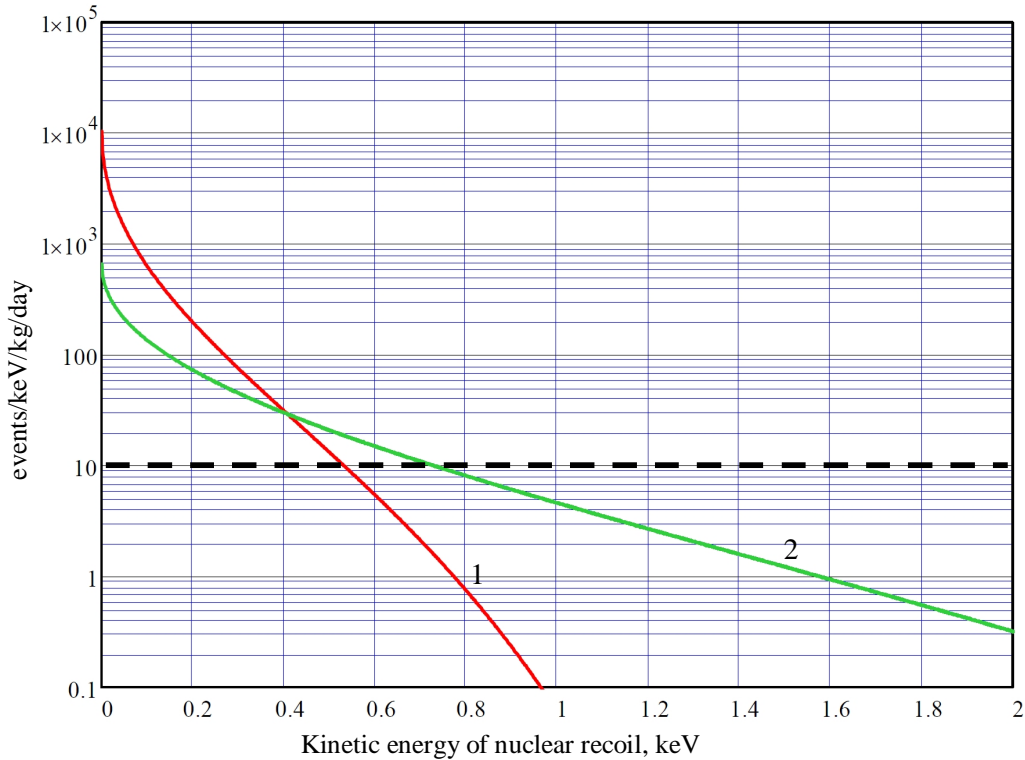

**Figure 1.** Energy distribution of nuclear recoils for the antineutrino flux of $10^{13}$ cm$^{-2}$s$^{-1}$. 1 – in liquid Xe, 2 – in liquid Ar. Dash line is a typical background level of Dark Matter detectors.

The obtained energy distribution of xenon (1) and argon (2) nuclear recoils is shown in figure 1. Calculations were performed for the neutrino flux of $10^{13}$ cm$^{-2}$c$^{-1}$. A typical background level of Dark Matter detectors is also shown by horizontal dash line[1]. To check the correctness of calculations, similar distribution was calculated for Si (not shown here), and we have found an absolute agreement with that obtained in [11]. Distribution of events in number of released ionisation electrons per event (shown in figure 2) is obtained from this distribution for several values of the ionisation yield: from 2 to 20 e$^-$/keV. The ionisation yield is not known precisely below 1 keV (for LXe), and in spite of the fact that extrapolation to this region gives with high probability the value of ~ 20 e$^-$/keV, the results are given for more pessimistic values as well. The expected number of events with different number of ionisation electrons is also given in table 1. A Poisson distribution of the number of released ionisation electrons is suggested for each ionization yield value, since a Fano factor is unknown for such low-energy depositions and it is assumed to be equal to 1. It is seen from the table that the number of events with at least one ionisation electron ($N_{e^-} \geq 1$) is significant even for the minimal ionisation yield. In the latter case, the results are not too far from the estimation made in [6],[7] with the use of SRIM code [12]. The number of events without ionisation may be slightly different from the real number in our case due to extrapolation of the antineutrino energy spectrum down to the region below 2 MeV using the formula (4) (see above). The only events with two electrons or more *will be of our interest*, because of the internal background of spontaneous single electrons which will prevail in a real detector (see below). It is seen that for this case ($N_{e^-} \geq 2$) the number of events is also significant even for the minimal considered value of the ionisation yield.

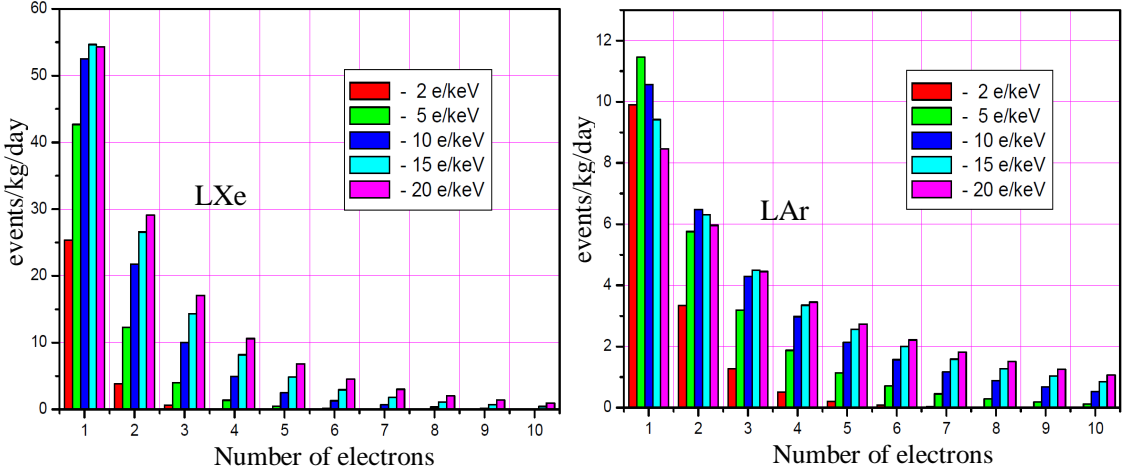

**Figure 2.** Distribution of events in number of electrons for liquid Xe (left) and liquid Ar (right). The estimations are made for antineutrino flux of $10^{13}$ cm$^{-2}$s$^{-1}$.

| Ionization yield (e$^-$/keV) | 2 | | 5 | | 10 | | 15 | | 20 | |
|---|---|---|---|---|---|---|---|---|---|---|
| Target | Xe | Ar | Xe | Ar | Xe | Ar | Xe | Ar | Xe | Ar |
| events with $N_{e^-} = 0$ (kg$^{-1}$day$^{-1}$) | 224 | 38 | 193 | 28 | 160 | 20 | 138 | 16 | 122 | 14 |
| events with $N_{e^-} \geq 1$ (kg$^{-1}$day$^{-1}$) | 30 | 15 | 61 | 25 | 94 | 33 | 116 | 37 | 132 | 39 |
| events with $N_{e^-} \geq 2$ (kg$^{-1}$day$^{-1}$) | 5 | 5 | 18 | 14 | 42 | 23 | 62 | 28 | 78 | 31 |

**Table 1.** Number of events with different number of ionisation electrons released in the detector per event for different values of the ionisation yield. These values are assessed for the antineutrino flux of $10^{13}$ cm$^{-2}$c$^{-1}$.

---
[1] The real energy scales for nuclear recoils and gamma-rays are different for the condensed noble gas media. In the nuclear recoil equivalent energy scale (taking into account a so-called quenching factor), this level of background is even lower.

In compare to the hypothetic Dark Matter particle (WIMP), the reactor antineutrinos produce much less energetic nuclear recoils. Consequently, detection of scintillation from them is practically impossible, and the detector must have the superior detecting performance for the very low ionisation signals.

## 3. Detector conception

A detector for this task should be developed in accordance to the following basic principles:
- it must detect extremely low ionisation, down to single ionisation electron;
- must possess good single electron resolution: the signal from one electron must be reliably distinguishable from the two-electron signal;
- must possess a coordinate resolution for the single electron events, not worse then few cm;
- must be able to discriminate background gamma and neutron events in the keV-energy range and to measure their energy spectrums with the aim of extrapolation of them to the range where the signal is expected;
- must be surrounded with a passive shield from gamma quanta and neutrons and an active shield as well from neutrons and cosmic muons;
- must have long-term stability and capability of at least one year operation;
- must be compact: no more then several cubic meters of the whole setup volume including the shield.

It is proposed to use a detector based on condensed noble gas with a two-phase technique of signal detection. The unique feature of such detector is a combination of good detecting properties of the liquid phase and amplification properties of the gaseous one. The liquid phase is a medium which possess the high density and the excellent transport properties for the free electrons. Ionisation produced by nuclear recoils from antineutrino scattering with a magnitude of several electron-ion pairs (according to the estimations shown above) will be detected by means of extraction of the electrons by the electric field to the gas phase. In the gas phase, the electrons produce the strong electroluminescent signal which is detected by ultraviolet sensitive photomultipliers. This method allows one to measure the number of ionisation electrons by counting them (see below). In order to do this, the electric field in the gas phase should be just below the threshold of avalanche multiplication to ensure only the high electroluminescence but without gas amplification. This is necessary for obtaining the Gaussian type of the single electron peak distribution with the smallest width.

## 4. Possible detector versions

Below are two versions of the detector (figure 3) based on the up-to-date technological achievements.

The first one (figure 3a) is close to that of the Dark Matter detector Xenon10. The detector sensitive volume with a height of ~15 cm and a diameter of ~20 cm is defined by a cylinder of field-shaping rings manufactured from an oxygen free copper. This volume contains ~15 kg of LXe or ~ 6.5 kg of liquid Ar (LAr). On the bottom of the volume, there is a flat grid cathode. The electric field of ~0.7 kV/cm is applied to the sensitive volume. On the top of the volume there are two grids placed at a distance of 12 mm from each other which provide the electric field necessary for extraction of the electrons from the liquid to gas phase (~4 – 5 kV/cm in the liquid) and for the electroluminescence (proportional scintillations) in the gas phase (~8 – 10 kV/cm). The thickness of the liquid above the lower of these two grids is 2 mm and the thickness of the electroluminescence gap (between the surface of the liquid and the upper anode grid) is 10 cm. The anode is maintained at zero potential, the lower grid is at -10 kV, and the cathode is at -20 kV. For screening of the photocathodes having a negative potential of ~ 1kV from the field leaking above the anode it is necessary to add one more grid maintained at the same potential. Therefore, the distance between the photocathode plane and the middle of the electroluminescent gap is about 1.5 cm that provides the good collection of the electroluminescent light. It is proposed to use the PMTs of the same type as in the Xenon10 detector: Hamamatsu R8520-AL in amount of 52. These photomultipliers with the high quantum efficiency (~ 30% in the emission range of Xe (175 nm)) were specially developed for operation in the liquid xenon medium. These PMTs are ultra low-radioactive: concentration of U, Th and K is, correspondently, 0.17±0.04, 0.20±0.09, 10±1 µBq per PMT. The advantage of these photomultipliers is also their square shape (2.5 x 2.5 cm$^2$ for the PMT and 2.2 x 2.2 cm$^2$ for the photosensitive area) that allows one the maximum packaging. For the detector filled with LAr a wavelength shifter coating of the PMTs could be used as it is done in the WARP Dark Matter detector [13]. Another possible option is to add small amount of xenon to argon to shift

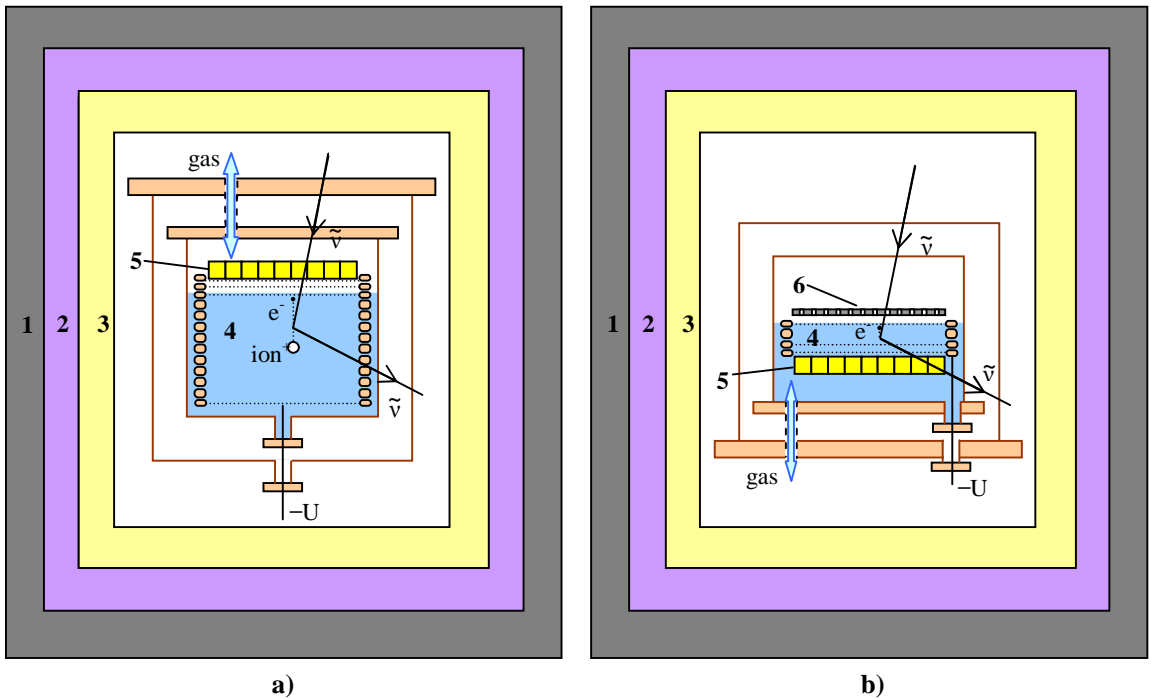

a)                            b)

Figure 3. Two versions of the conceptual design of the condensed noble gas-based two-phase detector: a) with the use of PMT, b) with the use of PMT and a GEM-like structure. 1 – lead shield, 2 – plastic scintillator, 3 – Gd-loaded plastic, 4 – liquid noble gas, 5 – 52 PMTs, 6 – GEM. *Out of scale.*

emission spectrum to 175 nm [14],[15].

In the second version (figure 3b), the photomultipliers are placed below the grid cathode, and they view the sensitive volume of the detector from the bottom. The PMTs in the Dark Matter detector ZEPLIN III are located in a similar way. The thickness of the sensitive volume should be less in this case (let's say, ~ 5 – 10 cm), and the cathode potential could be reduced, correspondingly. The amount of LXe and LAr in the sensitive region for this version is about 7.5 and 3 kg, correspondingly.

A Gas Electron Multiplier (GEM) [16] or thick GEM (THGEM) [17] structure is used in the place of the anode, in saturated vapour above the liquid [18],[19]. This structure will allow one to localize the event in the horizontal plane with an accuracy of ~ 1 mm. The particular technical realization of this structure is not under consideration here. One just accepts as a fact that it will have the gain sufficient for detection of single electrons (see for example [20]) and will be capable to measure their coordinates with an accuracy of few millimetres.

In both cases the detector will operate in a mode of discrimination of gamma and neutron background events, as it is done in Dark Matter detectors. Consequently, it will be possible to measure separately the backgrounds of both types and to extrapolate them to the energy range below 1 keV.

The detector is placed in a multilayer shield (see figure 3) consisted from lead (1; a gamma shield), plastic scintillator (2), and Gd-loaded hydrogen enriched plastic (3). Thickness of each layer is about 10 – 15 cm. The hydrogen enriched plastic serves for neutron moderation. The plastic scintillator operates in anticoincidence with the detector. The moderated neutrons are captured by Cd. The latter emits several gamma-rays (with a total energy of ~ 8 MeV) which are detected by plastic scintillator. For the neutrons coming from outside the layers 2 and 3 works like a passive shield which moderates the fast neutrons to such low energies that they either stopped in the Gd-loaded plastic or plastic scintillator, or produce the nuclear recoils in the detector with the energies much lower the detector threshold. In the case when the neutron has nevertheless passed through the shield (for example, through the "holes" in the shield) and produced the nuclear recoil in the energy region of interest, the scattered neutron will be moderated then and will produce the "veto" signal in the scintillation detector.

## 5. Single-electron characteristics

For detection of coherent scattering of reactor antineutrino the detector must operate in a single-electron (SE) counting mode. Therefore it is important to know its characteristics related to single-electron detection: pulse height distributions of the SE, time parameters of the signals.

The number of photons emitted by electron drifting in a uniform electric field above the liquid is an important characteristic of the detector. This number can be estimate with the following formula [21],[22],[23]:

$$N_{ph} = a \times x \times (E/n) + b \times x, \qquad (4)$$

where $E$ – electric field strength in the gas phase, $n = N_A \cdot r/A$ – density of atoms, $r$ – saturated vapour density, $A$ – average atomic number of the gas, $x = n \cdot d$ – number of atoms per unit of area for the electroluminescence gap thickness $d$.

For the xenon filled detector one can take $a = 0.137$ V$^{-1}$, $b = -4.7 \cdot 10^{-18}$ cm$^2$/atom, obtained in the work [21] for the saturated xenon vapour at cryogenic temperatures. Note, that practically the same values have been measured for xenon but at room temperatures in [22]. For $d$ = 1 cm, $E$ = 9000 V/cm, $r$ = 1.17·10$^{-2}$ g/cm$^3$ ([24]; at $p$ = 1.2 bar, a typical operation pressure of xenon two-phase detector) the number of photons will be equal $N_{ph}$ = 983.

For the argon filled detector there are no experimental data for a saturation vapour, and we may take the values obtained in [23] for the room temperature argon: $a$ = 0.081 V$^{-1}$, $b$ = – 1.9·10$^{-18}$ cm$^2$/atom. For the same operation electric filed and pressure ($r$ = 0.674·10$^{-2}$ g/cm$^3$ at $p$ = 1.2 bar [24]) the number of photons will be equal $N_{ph}$ = 536.

One should take into account that only 50% of photons are going in direction of PMTs, and the PMT geometrical efficiency is about 70% (ratio of the total photosensitive area to the total area of the PMT plane). For the Xe-filled detector (175 nm UV photons) we can take the photocathode quantum efficiency of 30%, therefore the single-electron signal from the PMT plane can be estimated as ~ 100 photoelectrons. For the Ar detector (125 nm UV photons) with 72% and 50% efficiency of TPB wavelength shifter ([25] and [26], correspondently) the single-electron signal will be equal to ~ 40 and ~ 30 photoelectrons. In reality, one should take the values by a factor of 2 lower, i.e. ~ 50 and ~ 20 photoelectrons for Xe and Ar, correspondently, since the significant part of the light comes to the PMT surface at large angles with respect to the normal, and therefore, undergoes significant reflection. In the Xenon10 detector, this value was 24 ± 7 photoelectrons [9] for $d$ = 0.5 cm that is in good agreement with our estimation.

The time structure of the signal is the following: it consists of separate single photoelectron signals going in an accidental order with roughly equal probability within the time interval of ~ 2 µs that corresponds to the total drift time of the electron in the 1-cm thick electroluminescent gap. The integrated area distribution of the single electron signals must be well described by the Poisson distribution, because an avalanche process which destroys the Poisson statistics does not start at such electric field values.

As regards the single-electron characteristics of GEM/THGEM multipliers, one should refer to the works [19],[20], where the efficient performances of the triple-GEM and double-THGEM multipliers in single-electron and two-electron counting modes, respectively, were demonstrated in two-phase Ar, at gains exceeding 5000 and 2000, respectively. It should be noticed that the electron collection efficiency into the GEM holes might decrease at higher drift fields (in front of the first GEM), required for efficient proportional scintillations and high electron extraction yield from the liquid. It would not be actually a problem for Ar detectors, since the electron collection efficiency in THGEMs could be equal to 100% right up to the fields of 5 kV/cm [17]. In Ar detectors, such fields are sufficiently high for the efficient scintillation and electron extraction from the liquid. In Xe detectors, however, it is necessary to operate at 9 kV/cm, which might result in the reduced collection efficiency, by a factor of 2 [17].

## 6. Background

Estimations of radioactive background in the energy range of interest have been made by authors of the work [7] for a two-phase argon detector. It has been shown that the counting rate of single-, two-, etc. electron events caused by radioactive background will be reduced down to the level of one to few events kg$^{-1}$ day$^{-1}$ after installation of the multilayer shield of the type shown in figure 3. This is close to the radioactive background level of Dark Matter detectors and significantly lower than the expected effect level.

However, the tests performed with a two-phase detector in ITEP have shown that the prevailing background in the experiment will be an apparatus background of spontaneous single-electron signals. The test with a trigger specially tuned on the single-electron signals with a characteristic sequence of single photoelectron signals with ~ 1-µs duration has shown that the rate of such events ($R$) in a typical ground-level laboratory conditions (without any additional shield from the external natural radioactivity) is about $0.5 \cdot 10^2$ Hz·kg$^{-1}$ (or ~ $4.3 \cdot 10^6$ kg$^{-1}$·day$^{-1}$). This rate is extraordinary higher than the expected effect rate. Such signals may have different origin. For example, one of the mechanisms may be a photoionisation which takes place on residual contaminations in xenon. As a rule, the single-electron signals are observed during several (or even tens) microseconds after quite intensive light flashes of scintillation and electroluminescence (within the maximum drift time of the charge in the detector sensitive volume). These signals can be excluded by blocking the trigger for a certain time after the electroluminescent signals. However, as the test has shown, such single-electron signals appear with high frequency even without link with any foregoing light pulses.

A field emission of electrons from a wire cathode (from the parts, which produces the high electric field, such as sharp edges, uncontrollable roughness, etc.) and emission from the free surface of the liquid noble gas might be responsible for such signals. In the latter case, it might be induced by delayed emission of the electrons remained under the surface due to incomplete extraction of the ionisation electron cloud to the gas phase at the moment of its arrival to the surface (the extraction efficiency to the gas phase is not equal to 100%). These undersurface electrons are drifting slowly towards the detector peripheral regions and could be emitted in a gas phase at any moment. Also, electronegative ions formed by capture of electrons with electronegative contaminations may stay for long time at the surface and at some conditions may release the captured electrons. In the first case, the frequency of the single-electron events depends on the quality of the cathode, and consequently, it may be reduced in accordance to this requirement. In the second case, it depends on the global detector counting rate, and consequently, can be reduced by surrounding the detector with a passive shield.

Let's, however, make estimations in a pessimistic scenario, i.e. in suggestion that the counting rate won't be reduced (*let's assume also the same single electron background rate for the Ar detector*). In this case, one should select only the events having two or more ionisation electrons. It is important to note that, most probably, one will eventually have to accept such event signature anyway, since it is quite evident that the background of spontaneous electrons can hardly be reduced to the desirable level. For selection of such events (having two or more ionisation electrons), the detector must possess the good single electron resolution, otherwise the background contribution from the single electrons may be significant.

One can estimate a discrimination factor for the single-electron background when the threshold is set between one and two electrons in assumption of a Poisson distribution for the single-electron pulse spectrum shape. Distributions of signals for the expected number of photoelectrons 50 and 100 in single- and two-electron peaks, respectively, are shown in figure 4a (corresponds to Xe detector).  The similar distributions for the expected number of photoelectrons 20, 40 and 60 in single-, two-, and three-electron peaks, respectively, are shown in figure4b (corresponds to Ar detector). The rejection factor of $10^6$ will be achieved at thresholds of ~ 90 and ~ 45 photoelectrons for Xe and Ar detectors, respectively.

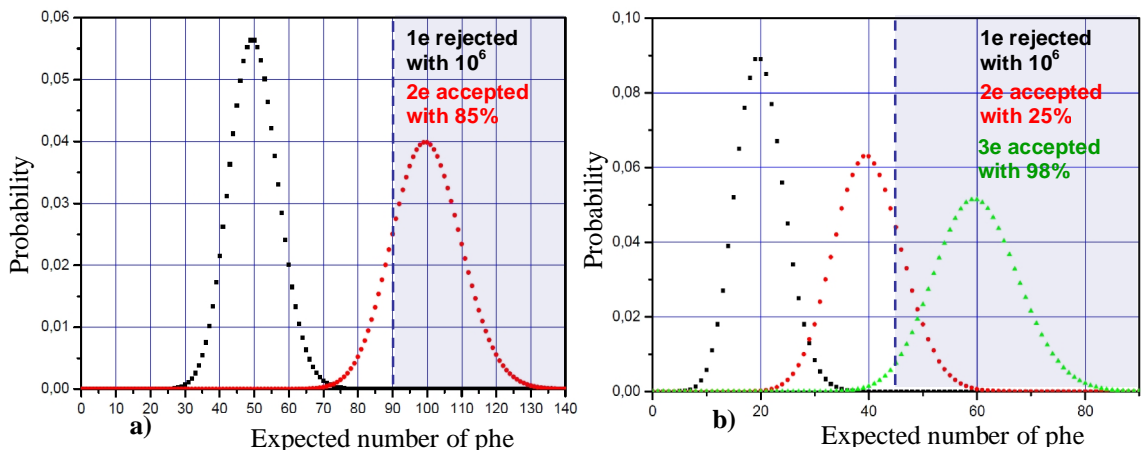

**Figure 4.** Poisson distributions for the expectations 50 and 100 (a) and 20, 40, and 60 (b).

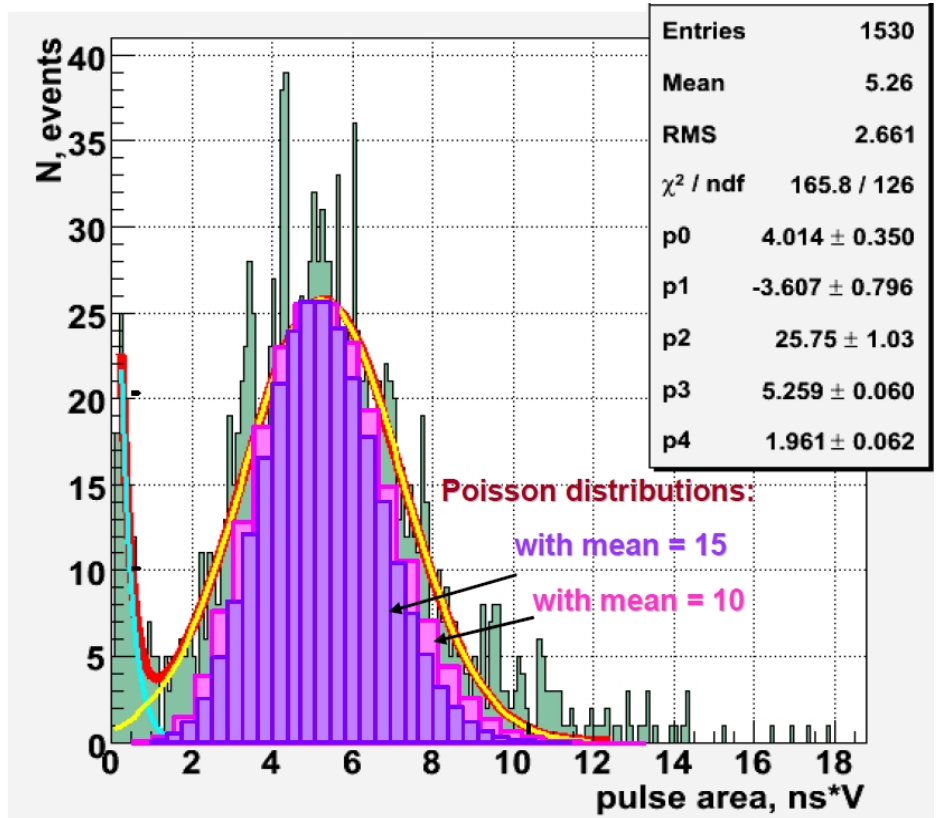

**Figure 5.** Experimental single-electron distribution, obtained with the ITEP two-phase Xe test detector [27]. The mean of Gaussian fit is 15±5 photoelectrons. Also, Poisson distributions with expectation 10 and 15 are shown.

At these trigger levels, the detection efficiency will be ~85% and ~ 0.25% for two electrons and ~100% and ~ 98% for three electrons for Xe and Ar, respectively. For the yield of ≥ 5 e⁻/keV, the low efficiency of two-electron events for Ar detector is not crucial because the probability of three- and more electron events is significant (see figure 2b). As a rule, the real distribution of the single-electron signal is a bit wider: see figure 5 which demonstrates a single-electron spectrum measured with a two-phase test detector in ITEP. The distribution is well fitted with a Gaussian with a mean value corresponding to 15±5 photoelectrons obtained from single photoelectron calibration [27]. Poisson distributions are shown by histograms in this figure for the expectations equal 10 and 15. It is clear that the width of the real distribution is wider than predicted one. Nevertheless, one may expect the discrimination factor of $10^6$ by the price of threshold increase, and as a consequence, by decreasing of the two-electron events detection efficiency, provided there is no non-Gaussian tails in the distributions. With

– 9 –

discrimination of $10^6$ the residual contribution from the single-electron background will be ~ 4.3 $kg^{-1} \cdot day^{-1}$ that is lower than the expected effect rate for the ionisation electron yield $\geq 2$ $e^-/keV$.

The spontaneous single-electronic signals may also coincide in time and mimic the two-electron ones. Let's admit that two electroluminescent signals from separate single electrons will be interpreted as one signal when the difference between their arriving times is $\leq \Delta T = 1$ µs, the value equal to one half of the signal duration (then, overlapping is $\geq 1$ µs, and the total signal duration is 2 – 3 µs). Then the frequency of such accidental coincidences is:

$$2 \times R^2 \times \Delta T = 5 \cdot 10^{-3} \text{ Hz} \cdot kg^{-1} = 432 \text{ } kg^{-1} \cdot day^{-1}.$$

The only method to discriminate this background is to use the capability of detector to separate the events with different X and Y coordinates (in the plane parallel to the anode): the real event having two or more ionisation electrons must be point-like. It is difficult to make a precise estimation of discrimination factor for this method now, because the spatial resolution for the single-electron events is unknown. For the first version of detector, we may take the spatial resolution (FWHM) of the same order of magnitude as a photocathode diameter. Then the discrimination factor can be estimated as a ratio of the whole PMT plane area to approximately four photocathode areas, i.e. ~ 10. The residual rate will be then 43 $kg^{-1} \cdot day^{-1}$ that is close to the expected effect rate for the specific electron yield $\geq 10$ $e^-/keV$.

In all probability, knowledge on reduction of the spontaneous single-electron background will be obtained in the studies of its origin, and the estimations given here may be considered as the upper ones. Note for example, that since the rate of accidental coincidences depends on R quadratically, reduction of R by a factor of 3 results in reduction of it by an order of magnitude.

For detector of the second version, the X,Y coordinates could be measured using a GEM/THGEM multiplier with an accuracy down to ~ 1 mm. This will allow one to suppress the accidental coincidences of single electrons by a factor of ~$10^4$, i.e. down to 0.0432 $kg^{-1} \cdot day^{-1}$. It is essential that the spectra in GEM/THGEM multipliers in two-phase Ar are exponential: see Figure 6 [20], which demonstrates a single-electron and 1.4-electron spectrum measured with the two-phase Ar avalanche detector in BINP. Nevertheless one can see that at higher gains, $\geq 10^4$, the single-electron spectrum is well separated from that of electronic noises, with Equivalent Noise Charge ENC=1000 e. The noise rate of the GEM structure itself can be estimated from the data of [19], where the triple-GEM was operated in two-phase Ar at a detection threshold of 4 primary electrons: the noise rate was as low as 0.2 Hz/cm$^2$ and thus can be neglected in a triggered mode.

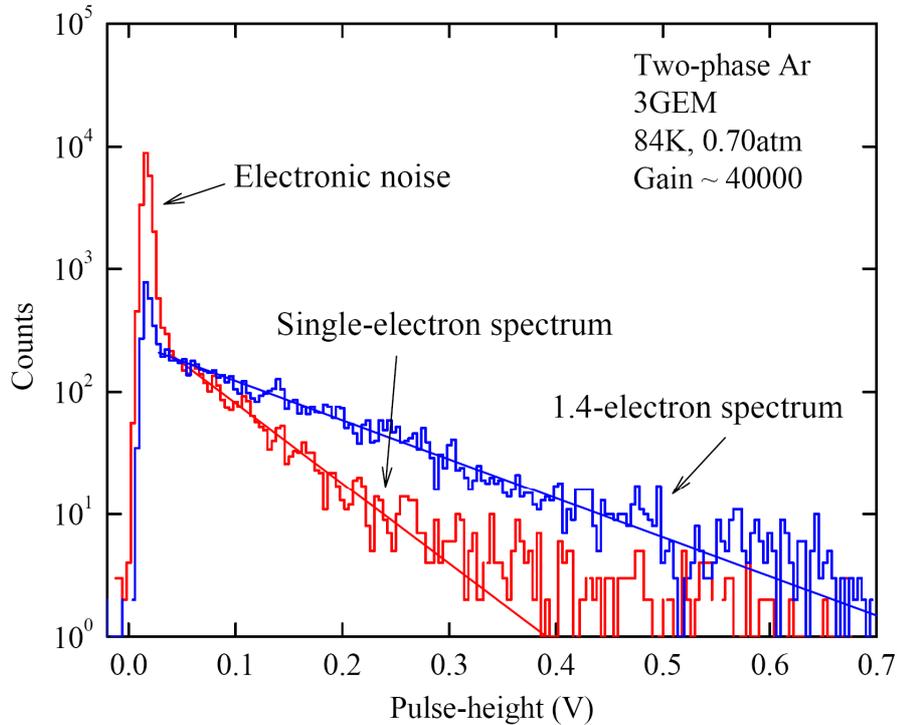

**Figure 6.** Single- and 1.4-electron distributions obtained with the two-phase Ar avalanche detector in BINP, using a triple-GEM multiplier at a gain of 40000 [20]. Electronic noise corresponds to ENC=1000 e.

## 7. Conclusion

Signal amplitudes and counting rates of the events of reactor antineutrino coherent scattering off atomic nucleus in a two-phase noble gas (Xe, Ar) detector are estimated on the basis of extrapolation of the recent experimental data on ionisation electron yield from the nuclear recoil track in liquid xenon [9] to the energy region < 1 keV.

The rather high frequency of spontaneous single-electron events in a two-phase detector observed in laboratory tests requires to select for triggering two or more electrons released in a single point. The estimations made both for Xe and Ar detectors for the neutrino flux of $10^{13}$ cm$^2$s$^{-1}$ show that with this selection the rate of the antineutrino scattering events will be from ~ 10 to ~ 300 kg$^{-1}$·day$^{-1}$ (varying for the different values of ionisation electron yield of $\geq 2$ e$^-$/keV ).

Two versions of detector are proposed. The first one utilizes electroluminescent amplification of the ionisation signal, and it *is totally based on detection techniques and technologies successfully used in the Dark Matter search experiments*. Building of such detector could be started immediately. The second one combines a power of very good single electron resolution of a first-version detector and a power of precise coordinate measurement by a gas electron multiplier (GEM or THGEM). However, some additional R&Ds are required on GEM performance in a saturated xenon vapour to achieve the necessary gain for observation and counting of single electrons. Nevertheless, it looks more attractive since the combined analysis of the independent information from PMTs and GEMs will allow one to identify the events with the best reliability.

## Acknowledgments

We thank A. Starostin for fruitful discussions. This work was supported by the Russian Foundation for Basic Research grant RFBR 05-02-08095-ofi_e.